\documentclass[aps,twocolumn,prl,groupedaddress,showpacs]{revtex4-1}
\usepackage[T1]{fontenc}
\usepackage[latin9]{inputenc}
\usepackage{times}
\usepackage{color}
\usepackage{xspace}
\usepackage{amssymb,amsmath}
\usepackage{amsbsy}
\usepackage{graphicx}
\usepackage{multirow}
\usepackage{rotating}
\usepackage{microtype}
\usepackage{tabularx}
\usepackage{epstopdf}
\usepackage{float}
\usepackage{mathrsfs}
\usepackage{hyperref}
\usepackage[normalem]{ulem}
\newcommand{\Niso}{$^{14}$N }
\def\be{\begin{equation}}
\def\ee{\end{equation}}
\def\e#1{\label{#1}\end{equation}}
\def\bea{\begin{eqnarray}}
\def\eea{\end{eqnarray}}
\def\nn{\nonumber}

\def\ket#1{{|#1\rangle}}
\def\bra#1{{\langle#1|}}

\begin{document}
\title{Dissipative entanglement of solid-state spins in diamond }
\author{D. D. Bhaktavatsala Rao }
\author{Sen Yang}
\author{J\"{o}rg Wrachtrup}

\affiliation{3. Physikalisches Institut, Research Center SCOPE, and MPI for Solid State Research, University of Stuttgart, Pfaffenwaldring 57, 70569 Stuttgart, Germany
}%
\date{\today}
\begin{abstract}
Generating robust entanglement among solid-state spins is key for applications in quantum information processing and precision sensing. We show here a dissipative approach to generate such entanglement among the hyperfine coupled electron nuclear spins using the rapid optical decay of electronic excited states. The combined dark state interference effects of the optical and microwave driving fields in the presence of spontaneous emission from the short-lived excited state leads to a dissipative formation of an entangled steady state. We show that the dissipative entanglement is generated for any initial state conditions of the spins and is resilient to external field fluctuations. We analyze the scheme both for continuous and pulsed driving fields in the presence of realistic noise sources.
 \end{abstract}
\maketitle
Originally inhomogeneities, decoherence and decay of quantum states were minimized in quantum computing proposals so that their effects would not disturb the ideal unitary evolution of the system \cite{book}. Recent works, however, suggest a quite opposite strategy where dissipation is used as a resource where  the system is driven on resonance with short lived states such that it dephases and decays to robust steady states \cite{kraus}. By suitable use of the interactions, these states can be selected, e.g., as entangled states or states encoding the outcome of a quantum computation \cite{dp1,dp2,dp3}. The remarkable feature of dissipative approaches is their resilience to errors that occur with  imperfect state initialization, to fluctuations in the driving field strengths and to dependencies on the system size - errors which on the other hand are quite harmful for unitary approaches employed to create the same entangled state \cite{unitary}. Dissipative approaches have been proposed to create entanglement among atoms or ions \cite{durga, soren, plenio, wine} and the robustness of dissipatively driven entanglement has been verified experimentally in ion traps \cite{wineland} and superconducting circuits \cite{sc}, and in the collective spin degrees of freedom of large atomic ensembles \cite{polzik}. While most of these studies used trapped atoms/ions as the physical system, we show here how the decay of electronic excited states in solids can also be exploited to generate deterministic entanglement among solid-state spins in diamond.

The negatively charged nitrogen-vacancy (NV) centers in high-purity diamond have been considered as a promising candidate for solid state quantum information processing due to their long coherence time and high feasibility in initialization, control, and readout of their spin states \cite{markus}. The NV center also provides a hybrid spin system in which electron spins are used for fast high-fidelity control \cite{acontrol,wcontrol} and readout \cite{nread, eread}, and the hyperfine coupled nuclear spins to store quantum information due to their ultra-long coherence time \cite{lukin}. Electron and nuclear spins could form a small-scale quantum register  allowing for e.g. necessary high-fidelity quantum error correction \cite{wcontrol,hec}. For experiments at both the room ($300K$) and low-temperatures ($4K$) the equilibrium states of electron and nuclear spins are close to a fully mixed state. Hence, any quantum protocol implemented using these spins needs a proper initialization step. Both theoretical and experimental methods to entangle the electron and nuclear spins follow the standard unitary scheme where specific initialization and timing of quantum gates is required to entangle the solid-state spins \cite{entang}. Here we alleviate the problem by taking a non-unitary approach to generate a steady state  entanglement of electron and nuclear spins for any initial state conditions. The optical excitation process usually used to only polarize the electron spins in diamond \cite{hanson} will be used to polarize the coupled electron-nuclear spin system into a maximally entangled state.

We start with an optically detectable single NV center consisting of an electronic spin (S=1) and intrinsic \Niso nuclear spin (I=1), coupled by hyperfine interaction. The Rabi driving between the ground state levels of the electron and nuclear spins can be  achieved by applying microwave (MW) and radio-frequency (RF) fields respectively as shown in Fig.1. In addition the electron can also be optically pumped into the excited state $\ket{A_1}$ that has a very short life time, and decays into all three ground states with different branching ratios. The Hamiltonian for the electron-nuclear spin system subject to various driving fields and interaction is given by
\bea
H &=& \Omega_e S_x+ \Omega_n I_x +gS^zI^z \nn \\
&+& (E_+\ket{+1}_e \bra{A_1} +E_-\ket{-1}_e \bra{A_1}+h.c) 
\eea
where the electronic and nuclear spin operators are respectively given by $S_x= (\ket{0}_e \bra{+1}_e + \ket{0}_e \bra{-1}_e + h.c)$ , $S_z= (\ket{+1}_e \bra{+1}_e - \ket{-1}_e \bra{-1}_e )$, and $I_x =(\ket{0}_n \bra{+1}_n + \ket{0}_n \bra{-1}_n + h.c)$,  $I_z= (\ket{+1}_n \bra{+1}_n - \ket{-1}_n \bra{-1}_n )$. The electron and nuclear Rabi fields are denoted by $\Omega_e$ and $\Omega_n$ respectively and their hyperfine interaction by $g$. The strength of the polarization dependent excitation of the electronic ground state levels $\ket{+1}_e$ and $\ket{-1}_e$ to $\ket{A_1}$ are denoted by $E_\pm$ respectively. Due to a very short life time the excited state $\ket{A_1}$ decays rapidly into the three ground states $\ket{\pm1}_e$ and $\ket{0}_e$ with decay rates $\gamma_{\pm}$ and $\gamma_0$ receptively.

\begin{figure}
\hspace{-5mm}
\includegraphics[width=80mm,height=50mm]{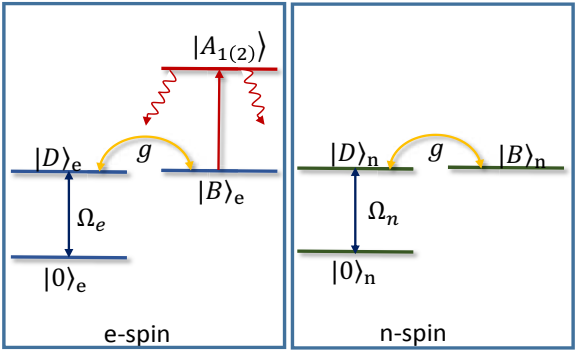}
\label{setup}
\caption{(Color online) Schematic representation of the relevant energy-level structure of the electron and nuclear spins written in dark, bright basis: $\ket{D}_{e/n} = \frac{1}{\sqrt{2}}(\ket{+1}_{e/n}+\ket{-1}_{e/n})$, $\ket{B}_{e/n} = \frac{1}{\sqrt{2}}(\ket{+1}_{e/n}-\ket{-1}_{e/n})$. The Rabi driving between the electron and nuclear spin ground states is generated by the microwave ($\Omega_e$) and radio-frequency ($\Omega_n$) fields respectively. The electron spins can be optically driven to the excited state $\ket{A_{1(2)}}$ from which it can rapidly decay into the ground states. The hyperfine coupling between the electron and nuclear spins, $g$, allows for the coupling between the bright and dark states $\ket{B}, ~\ket{D}$ of both the spins. }
\end{figure}

To describe the main idea behind the dissipative scheme it is useful to analyze different components of the above Hamiltonian, and look for states that remain stationary during the evolution. For example the hyperfine interaction $gS^zI^z$ does not have any role when either of the electron or nuclear spins are in the zero magnetic state $\ket{0}_{e/n}$, and hence states like $\ket{0}_e\ket{\psi}_n$ and $\ket{\psi}_e\ket{0}_n$ do not evolve under the hyperfine interaction. For the driving fields $\Omega_e S_x+ \Omega_n I_x$, with $\Omega_e = \Omega_n$, the singlet combination 
\be
\ket{\Psi}_{D} = \frac{1}{\sqrt{2}}\left[\ket{D}_e\ket{0}_n - \ket{0}_e\ket{D}_n\right].
\ee
remains stationary, where $\ket{D}_{e/n} = \frac{1}{\sqrt{2}}(\ket{+1}_{e/n}+\ket{-1}_{e/n})$. This state also remains stationary under hyperfine coupling as explained above. Now we would like to know under what conditions the above state could also remain stationary to the optical fields. For this we look into the selection rules for optical excitation from $\ket{\pm 1}_e$ to $\ket{A_1}$. Following \cite{adam}, one can shown that the optical $\Lambda$-system, with $E_+ = E_-$, has a dark state $\frac{1}{\sqrt{2}}(\ket{+1}_e+\ket{-1}_e)$, i.e., if the ground state of the electron spin is $\frac{1}{\sqrt{2}}(\ket{+1}_e+\ket{-1}_e)$, then it does not get excited. Using this result we can see that the state$\ket{\Psi}_{D} $ also remains stationary under optical driving. Remarkably, we have identified a state that remains dark under the unitary evolution of the system, and this state is the maximally entangled state of the electron and nuclear spins. Once the system is pumped into this dark entangled state, it becomes the long-lived state as it commutes with all the components of the Hamiltonian and their combinations.

\begin{figure}
\includegraphics[width=96mm]{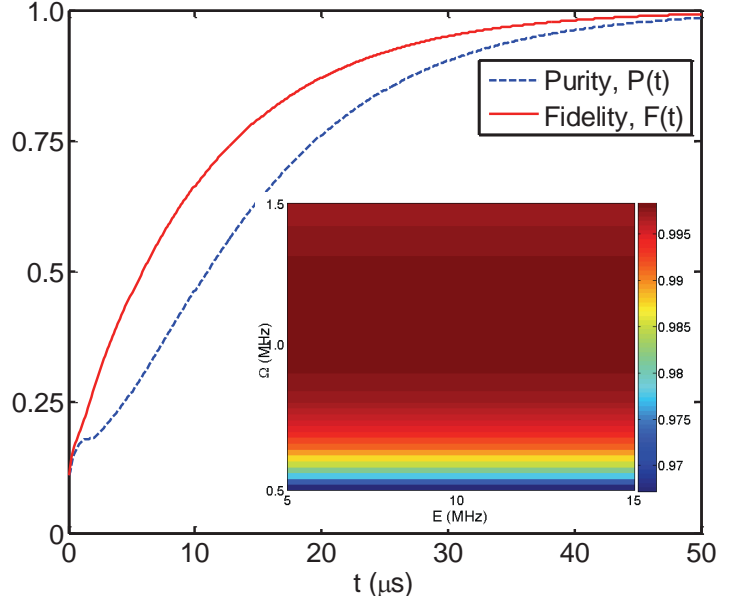}
\label{setup1}
\caption{(Color online) The fidelity $F(t)=\bra{\Psi}_{D}\rho_{en}(t)\ket{\Psi}_{D}$ (red solid-line), the purity of the electron-nuclear spin state $P(t) = Tr[\rho^2_{en}]$ (blue dashed-line) are plotted as functions time. The parameters chosen for the calculation are $\Omega_e = \Omega_n = 1$ MHz,$g = 2.5$MHz, $E_{\pm}= 10$MHz, and the decay rates $\gamma_{\pm}=30$MHz  MHz,$\gamma_{0}=40$MHz. The initial state is a fully mixed state in the ground state basis of the electron and nuclear spins. In the inset we show the purity of the coupled electron-nuclear spin system for various field strengths $E$ and $\Omega$.}
\end{figure}

\begin{figure}
\includegraphics[width=90mm]{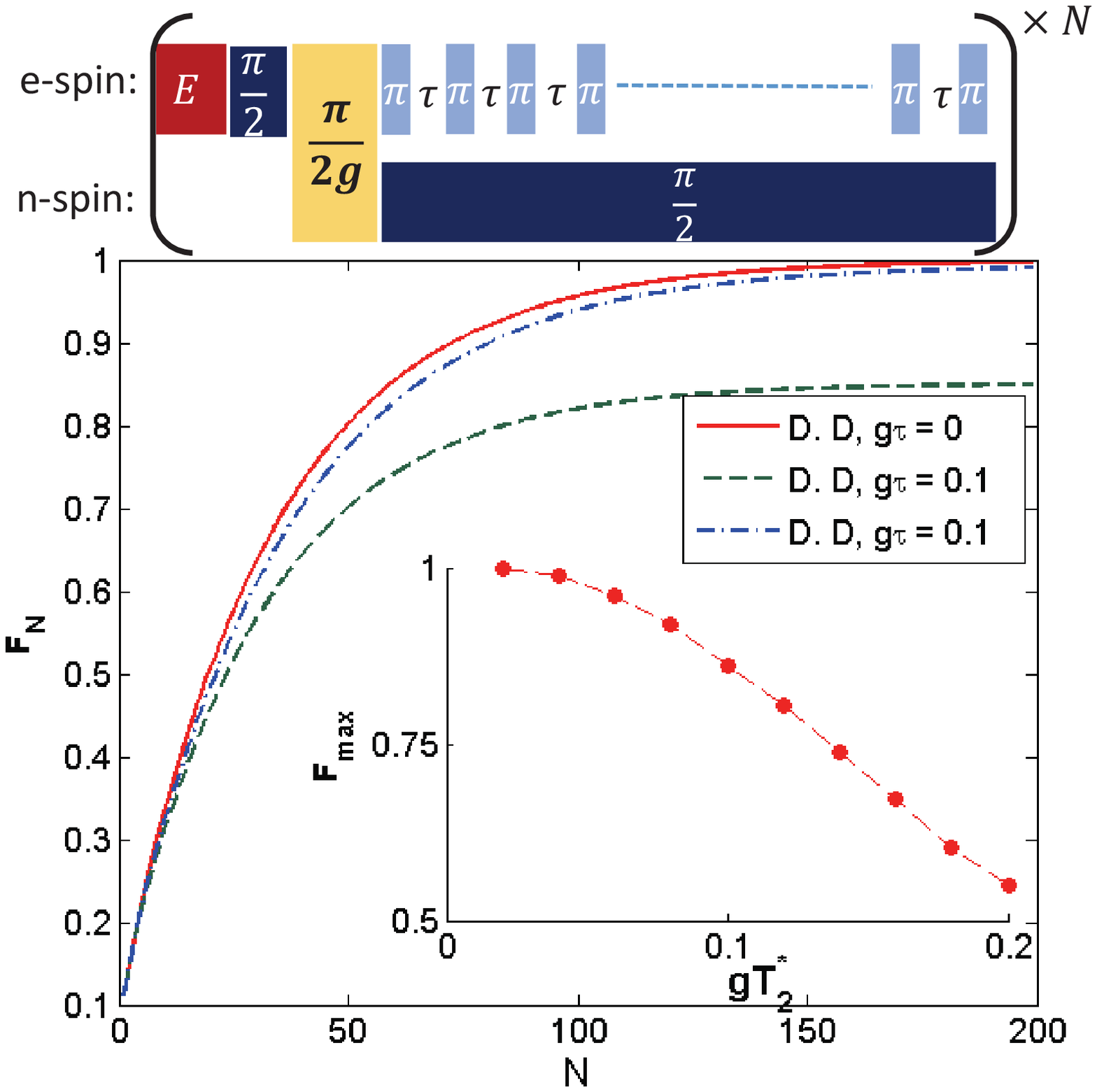}
\label{setup1}
\caption{(Color online) (Top) Schematic illustration of the pulses that are applied on the electron and nuclear spins. (Bottom) The fidelity is plotted as a function of the optical cycles $N$, for cases when the dynamical decoupling (D. D) sequence on the electron spins is exact (red solid-line), imperfect D. D (green dashed-line) that cannot filter certain low frequency noise ($\tau>0$) and the improvement observed by correcting the electron spin rotation in the case the imperfect D. D (blue dot-dashed line). In the inset we show the dependence of the maximal fidelity on the electron spin decoherence $T_2^*$.   }
\end{figure}
Clearly, the dark entangled state cannot be prepared by the unitary dynamics alone as it commutes with the Hamiltonian. Hence, one has to rely on the non-unitary dynamics driven by the strong dissipation of the excited state $\ket{A_1}$.
The theorem describing the dissipative generation of a many-body system \cite{kraus} states that one can identify a set of Lindblad dissipation operators that will turn a (zero) eigenvalue eigenstate of a Hamiltonian into the unique dark, steady state of the dissipative dynamics of the system. Though generation of such a set is in general a non-trivial task, our system offers a straightforward solution as the dark-state $\ket{\Psi}_D$ does not have any excited state components or those which could be excited to the unstable excited state. To verify, our claim we solve the dynamics generated by the ($12 \times 12$) master equation of the total system exactly. The state of the total system at any time can be obtained from 
\be
\partial_t \rho_{en}(t) = -i(\mathcal{H}\rho_{en}-\rho_{en}\mathcal{H}^\dagger) + \sum_{k} \mathcal{C}_k\rho_{en}\mathcal{C}^\dagger_k, ~~\mathcal{H} = {H}-\frac{i}{2}\sum_{k}\mathcal{C}^\dagger_k\mathcal{C}_k
\ee
where $\mathcal{C}_k = \sqrt{\gamma_k}\ket{k}_e\bra{A_1}$ are Lindblad operators, which describe the decay to the $k^{th}$ ($k = -1, +1, 0$) ground state of the electron. As $\ket{A_1}$ decays with equal probability to $\ket{\pm 1}_e$ states we can set $\gamma_+ = \gamma_-$. We would like to note that this condition is not necessary to obtain the dark state $\ket{\Psi}_D$. 

In Fig. 2 we have plotted the fidelity $F(t)=\bra{\Psi}_D\rho_{en}(t)\ket{\Psi}_D$, and the purity $P(t)=$Tr($\rho_{en}(t)^2$) of the state determined by solution of the master equation given above, starting from the fully mixed state of the electron and nuclear spins $\rho_{en}(0) = \rho_e(0)\otimes\rho_n(0)$ (the results are similar for any other initial state), where $\rho_{e/n}(0) = \frac{1}{3}[\ket{+1}_{e/n}\bra{+1}+\ket{-1}_{e/n}\bra{-1} + \ket{0}_{e/n}\bra{0}]$. The results confirm the evolution described above, where the initial mixed state with the lowest value of purity (1/9), and a very low entanglement fidelity (1/12) evolves towards the steady state $\ket{\Psi}_{D}\bra{\Psi}_{D}$ that that has unit purity and entanglement fidelity. The rate of entanglement generation is independent of the strength of the applied fields and the coupling $g$. These parameters affect only a change in the rate at which one can obtain the steady state given in Eq. (2). We show this behavior in the inset of Fig. 2 where over a broad range of applied fields the fidelity remains almost identical.

 As the electron and nuclear spin are inevitably coupled to a spin-bath comprised of the surrounding ${}^{13}C$ nuclear spins, they lead to random couplings between the dark ($\ket{D}_{e/n})$ and bright states ($\ket{B}_{e/n})$ of the electron and nuclear spins. Due to this coupling to the spin bath  $\ket{\Psi}_{D}$ is no more the stationary state for the dynamics. In addition to the spin noise errors, another major source for errors could arise due to the large mismatch in the Larmor frequency of these spins when equal powers of MW and RF fields are used i.e., when $\Omega_e \ne \Omega_n$. To overcome this mismatch one can use either a low MW field for driving the the electron or a very high RF field for the nuclear spin. While in the former case, a low MW field would allow the accumulation of spin noise errors, in the latter, high RF powers leads could lead to thermal noise due to heating of the RF components. To alleviate these problems we switch to the pulsed implementation so as to take into account both the slow manipulation of the nuclear spins and also the active filtering of spin noise by dynamical decoupling sequences applied on the electron spin, as shown in Fig. 3. As the strength of different fields and couplings involved in the entanglement generation are largely separated the total evolution can be decomposed into individual parts of electron and nuclear spin manipulations which we detail below. Similar to the continuous case discussed earlier where both the electron and nuclear spins are driven, the key requirement in the pulsed scheme is that  both the spins have to be equally rotated though not on a similar time scale.

The resonant optical excitation and the subsequent relaxation of the electron spins to the ground state is a rapid process and happens within few tens of nanoseconds. As there is minimum hyperfine coupling in the excited state $\ket{A_{1(2)}}$ this optical excitation can be treated completely independent and stroboscopic to nuclear spin effects. Following the optical excitation, a quick $\pi/2$ pulse is performed in the electron spin subspace spanned by $\ket{0}_e$ and $\ket{D}_e$. With the possibility to manipulate the electron spin with strong microwave fields, such a pulse can be performed within few nanoseconds. Thus, the electronic spin manipulation can also be treated stroboscopic. After this pulse we allow a free evolution where the electron and nuclear spins interact via their hyperfine coupling for an optimal time $\pi/2g$. As the hyperfine coupling strength is on the order of few $MHz$, there will be a finite noise accumulation over this time scale due to $T_2^*$ effects. Now we perform the nuclear spin $\pi/2$-pulse in the subspace spanned by $\ket{0}_n$, $\ket{D}_n$, and this could be an extremely slow operation in the $\sim 10$ microseconds. 

To achieve an ideal nuclear spin rotation it should remain decoupled to the electron spin during its manipulation. Since the RF fields used for nuclear spin rotations are much weaker than the hyperfine coupling, the electron spin should be dynamical decoupled from the nuclear spin by dynamical decoupling pulses (i.e., periodically flipping) at a rate $\tau \ll 1/g$. As $\tau$ can never reach zero, it leads to a finite error $\epsilon$ in the nuclear spin rotation, where, $\epsilon \approx \sin( \frac{g^2\tau}{\sqrt{g^2+\Omega^2_n}}$). This reduces the entanglement fidelity as we shown in Fig. 3. To correct this error and achieve maximal entanglement, we rotate the electron spin also by the same angle i.e., $\pi/(4+\epsilon)$. As we shown in Fig. 3 we regain the full entanglement after this correction. The errors due to $T_2^*$ mentioned above during the free evolution are shown in the inset of Fig. 3.

This scheme can be further extended to generate entanglement within a nuclear spin ensemble in addition to its entanglement with the electron spin. Such a spin ensemble composed of $13C$ spins (spin-$1/2$) is readily available in diamond and are coupled to the NV by hyperfine interaction. For example a nuclear spin ensemble of two spins and the NV which are initially in thermal equilibrium can be dissipatively driven into the steady state
\be
\ket{\psi_D} = \frac{1}{\sqrt{2}}\left[\ket{D}_e(\ket{10}_n + \ket{01}_n) +i \ket{0}_e(\ket{11}_n + \ket{00}_n)\right].
\ee
with $75\%$ while the remaining $25\%$is trapped in the nuclear spin-singlet state $\ket{10}_n - \ket{01}_n$. Due to the symmetry of the applied field singlet state remains unaffected and does not contribute to the dissipative process. To break this symmetry and to completely stabilize the system in the above state we switch on an asymmetry among the nuclear spins by manipulating them individually. 

The proposed scheme can be implemented efficiently at low temperatures (T$<$8 K) in a low strain ($\approx$1.2 GHz) NV center such that optical transitions, $A_{1(2)}$, are well resolved and there is no mixing between the levels  allowing for resonant excitation of the electron spin \cite{togan}. The degeneracy of the ground states can be maintained by switching off any external magnetic field. The coherence of the nuclear spin and the electron spin can be longer than a second at such low temperatures. The key source of error in the protocol is the $T_2^*$ time of the electron spin, and to achieve high fidelity $gT_2^* \ll 1$. While this noise can be filtered actively during the nuclear spin rotation, the error mainly accumulates during the free evolution time where the electron and nuclear spins evolve under the hyperfine coupling. With typical optical Rabi fields $\sim 30$ MHz, hyperfine coupling $\sim 2$ MHz, and the $\pi/2$-pulses of electron ($\sim 10$ ns), nuclear spin ($\sim 10~\mu$s)  a fidelity of $98 \%$ could be achieved in $\sim 2$ ms. This fidelity gets reduced to $95 \%$, for $T_2^* \sim 10~\mu$s.

In conclusion, we have shown that spontaneous emission of electron spins in diamond driven by resonant optical fields could assist not only in achieving ground state polarization but also to drive the coupled electron-nuclear spin system into a maximally entangled steady state. The generated entanglement can be kept alive beyond the $T_1$ times of the electron and nuclear spins, by keeping the external fields on. 
To scale up the dissipative approach to entangle multiple nuclear spins, an asymmetric nuclear spin manipulation would be required alongside the dissipation of the electron spins.  Dissipation could also play a key role in achieving entanglement between randomly oriented NV centers allowing it to be useful for applications in ensemble quantum sensing.

\begin{acknowledgements}
        We would like to acknowledge the financial support by the ERC project SQUTEC, DFG (FOR1693), DFG SFB/TR21, EU DIADEMS, SIQS, Max Planck Society and the Volkswagenstiftung.
\end{acknowledgements}

%
%
%
%
%

\end{document}